\documentclass[namedreferences]{solarphysics}
\usepackage[hyperref,optionalrh]{spr-sola-addons}  
\usepackage{graphicx}                                              
\usepackage{amssymb}                                            
\usepackage{color}                                                    
\usepackage{breakurl}                                              

\newcommand{\arcsec}{^{\prime\prime}}

\newcommand{\apj}{    {\it Astrophys. J.}}

\chardef\us=`\_

\begin{document}

\begin{article}
\begin{opening}
 
\title{Energy Partition in Four Confined Circular-Ribbon Flares}

\author[addressref={aff1,aff2},corref]{\inits{Z.M. Cai}\fnm{Z.M. Cai}}
\author[addressref={aff1,aff2,aff3},corref,email={zhangqm@pmo.ac.cn}]{\inits{Q.M. Zhang}\fnm{Q.M. Zhang}}
\author[addressref={aff1,aff2},corref]{\inits{Z.J. Ning}\fnm{Z.J. Ning}}
\author[addressref={aff1,aff2},corref]{\inits{Y.N. Su}\fnm{Y.N. Su}}
\author[addressref={aff1,aff2},corref]{\inits{H.S. Ji}\fnm{H.S. Ji}}
\address[id=aff1]{Key Laboratory of Dark Matter and Space Astronomy, Purple Mountain Observatory, CAS, Nanjing 210023, China}
\address[id=aff2]{School of Astronomy and Space Science, University of Science and Technology of China, Hefei 230026, China}
\address[id=aff3]{CAS Key Laboratory of Solar Activity, National Astronomical Observatories, CAS, Beijing 100101, China}

\runningauthor{Cai et al.}
\runningtitle{Energy Partition in Four Confined Circular-Ribbon Flares}

\begin{abstract}
In this study, we investigated the energy partition of four confined circular-ribbon flares (CRFs) near the solar disk center, which are observed simultaneously by SDO, GOES, and RHESSI.
We calculated different energy components, including the radiative outputs in 1$-$8, 1$-$70, and 70$-$370 {\AA}, 
total radiative loss, peak thermal energy derived from GOES and RHESSI, nonthermal energy in flare-accelerated electrons, and magnetic free energy before flares. 
It is found that the energy components increase systematically with the flare class, indicating that more energies are involved in larger flares. 
The magnetic free energies are larger than the nonthermal energies and radiative outputs of flares, which is consistent with the magnetic nature of flares. 
The ratio $\frac{E_{nth}}{E_{mag}}$ of the four flares, being 0.70$-$0.76, is considerably higher than that of eruptive flares. 
Hence, this ratio may serve as an important factor that discriminates confined and eruptive flares. 
The nonthermal energies are sufficient to provide the heating requirements including the peak thermal energy and radiative loss. 
Our findings impose constraint on theoretical models of confined CRFs and have potential implication for the space weather forecast. 
\end{abstract}
\end{opening}

\section{Introduction}  \label{s-intro}
Solar flares and coronal mass ejections (CMEs) are the most energetic activities in the solar system, which are considered as the main source of space weather \citep{fle11,webb12,gol16,pats20}.
The accumulated magnetic free energy (10$^{29}$$-$10$^{33}$ erg) in active regions (ARs) is released within a short period of time via magnetic 
reconnection \citep[e.g.,][]{ant99,amr00,chen00,lin00,mo01}. 
The released energy goes into the thermal energy of localized hot plasmas, kinetic energy of reconnection outflows, kinetic energy of CMEs, 
nonthermal energies of the accelerated electrons and/or ions, and radiations from radio to hard X-ray (HXR) 
and even $\gamma$-ray wavelengths \citep{for96,sto07,kre10,mill12,cas14,ing14,war16a,war16b}.
According to their association with CMEs, flares are classified into confined and eruptive types \citep[e.g.,][]{mo01,cheng11,su11,su15,ver15,li20,kli21}. 
A large number of confined flares result from failed filament eruptions due to the strong confinement of the overlying field \citep{ji03,liu14,sun15,zqm15,yang18,yan20}. 
Sometimes, confined flares are triggered by loop-loop interaction \citep{su13,kush14,ning18}.

Contrary to two-ribbon flares, circular-ribbon flares (CRFs) are a special type of flares that consist of a short, compact inner ribbon and a bright, outer ribbon with a circular or elliptical 
shape \citep{mas09,jos15,liu15,her17,devi20,kash20,pra20,jos21}. 
The three-dimensional (3D) magnetic configuration of CRFs is usually related to a fan-spine structure associated with a magnetic null point \citep{wang12,zqm12,sun13,hou19,lee20,liu20,yang20,zqm21}.
CRFs are occasionally accompanied by coronal jets or cool surges \citep{zqm16a,li17,xu17,dai20,zqm20}.
The dynamic evolution of CRFs, including magnetic reconnection near the null point, particle acceleration and precipitation, chromospheric evaporation and condensation, 
are found to resemble those of two-ribbon flares \citep{zqm16b,zqm19a}.

Till now, comprehensive investigations on energetics of eruptive flares are abundant \citep{ems04,ems05,ems12,mill14,war16a,war16b}. 
The energy partitions in flares and CMEs are comparable, especially for X-class eruptive flares \citep{feng13}.
Using multiwavelength observations from the \textit{Atmospheric Imaging Assembly} \citep[AIA;][]{lem12} on board the \textit{Solar Dynamics Observatory} (SDO), the energetics of 
nearly 400 eruptive flares were studied in detail \citep{asch14,asch15,asch16,asch17}. However, the investigation on energy partition in confined flares is rare. 
\citet{tha15} studied an X1.6 flare in AR 12192 on 2014 October 22. 
The nonthermal energy ($\sim$1.6$\times$10$^{32}$ erg) in flare-accelerated electrons is found to account for $\sim$10\% of the free magnetic energy before flare.
\citet{kush15} studied an M6.2 flare in AR 10646 on 2004 July 14. The time evolution of thermal energy is found to show a good correlation with the variations in cumulative nonthermal energy,
validating the well-known Neupert effect in confined flares.
\citet{zqm19b} explored various energy components in two homologous confined CRFs of the same class (M1.1) in AR 12434, including the peak thermal energy, nonthermal energy in electrons, 
total radiative loss of hot plasma, and radiative output in 1$-$8 {\AA} and 1$-$70 {\AA}. The two flares have similar energy partition, and the nonthermal energy is sufficient to 
provide the heating requirement incorporating the peak thermal energy and radiative loss.

In this study, we selected four confined CRFs near the solar disk center observed by SDO/AIA \citep{song18}. In Section~\ref{s-obs}, we briefly describe the data sets and calibration. 
We are not interested in the triggering mechanism of each flare, which has been extensively studied. We focus on the estimation of various energy components of flares in Section~\ref{s-eng}. 
The results are compared with previous findings in Section~\ref{s-dis}. Finally, a summary is given in Section~\ref{s-sum}.

\begin{table}
\caption{Information on the four flares in our study. 
Here, $t_{sta}$, $t_{peak}$, and $t_{end}$ represent the start, peak, and end times of the flares in GOES 1$-$8 {\AA}, respectively.
$\Delta t$ denotes the rough lifetime.}
\label{tab-1}
\tabcolsep 1.5mm
\begin{tabular}{ccccccccc}
  \hline
Flare & Date & $t_{sta}$ & $t_{peak}$ & $t_{end}$ & $\Delta t$ & AR & Location & Class\\
         &          & (UT) & (UT) & (UT) & (min) &  &  &  \\
  \hline
CRF1 & 10$-$May$-$2012 & 20:20 & 20:26 & 21:20 & 60 & 11476 &N12E12 & M1.7\\
CRF2 & 07$-$Nov$-$2013 & 03:34 & 03:39 & 03:54 & 20 & 11890 & S13E28 & M2.3\\
CRF3 & 29$-$Dec$-$2013 & 14:38 & 14:45 & 15:10 & 32 & 11936 & S16W05 & C5.1\\
CRF4 & 05$-$Mar$-$2014 & 00:10 & 00:15 & 00:25 & 15 & 11991 & S27W07 & C4.8\\
  \hline
\end{tabular}
\end{table}

\section{Data Sets and Calibration} \label{s-obs}
\begin{figure} 
\centerline{\includegraphics[width=1.\textwidth,clip=]{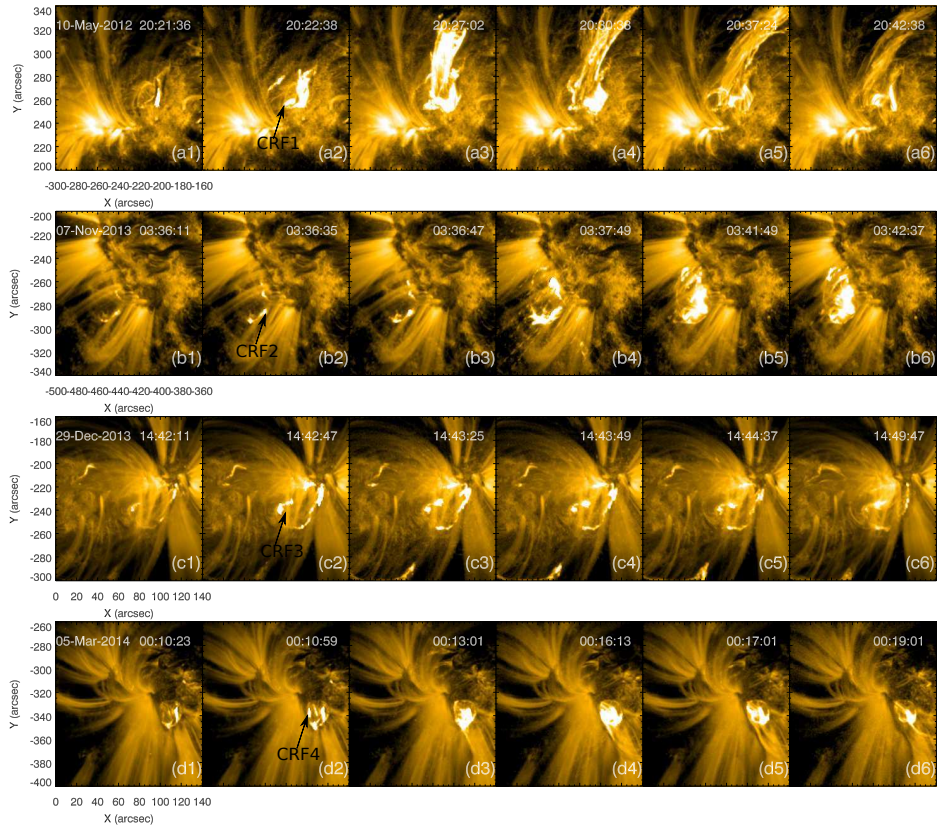}}
\caption{Snapshots of the four flares in AIA 171 {\AA}. The top row is for CRF1 and the bottom row is for CRF4. (Four animations of this figure are available online.)}
\label{fig1}
\end{figure}

The date/time, location, and GOES class of the four flares (CRF1, CRF2, CRF3, and CRF4) are listed in Table~\ref{tab-1}. Among the four flares, two are M-class and the others are C-class.
The flares were observed by SDO/AIA in extreme-ultraviolet (EUV) wavelengths (131, 171, and 304 {\AA}).
The AIA level\_1 data were calibrated using the standard Solar Software (SSW) program \texttt{aia\_prep.pro}.
In Figure~\ref{fig1}, the 171 {\AA} images illustrate the whole evolution of the flares (see also the online movies).
The four flares share a basic similarity in evolution. The inner ribbons and part of the outer ribbons brightened first. 
Then, the rest of the outer ribbons brightened sequentially, which is consistent with previous findings \citep{li17,xu17}. 
Finally, the brightness of flare ribbons declined gradually with time and died out. 
It is noted that CRF1 was associated with a blowout coronal jet propagating in the northwest direction, while the remaining three flares were not associated with jets. 

\begin{figure}
\centerline{\includegraphics[width=0.75\textwidth,clip=]{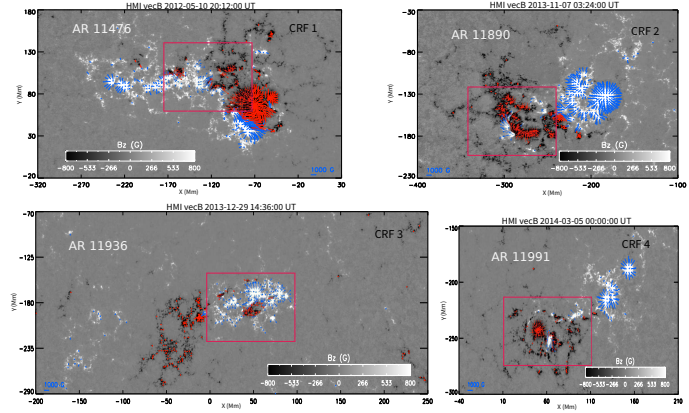}}
\caption{HMI vector magetograms of the four ARs hosting the flares.
NLFFF extrapolations were performed using these magnetograms.
The red boxes indicate regions for calculating the magnetic free energy, where flares took place.}
\label{fig2}
\end{figure}

\begin{figure}
\centerline{\includegraphics[width=0.75\textwidth,clip=]{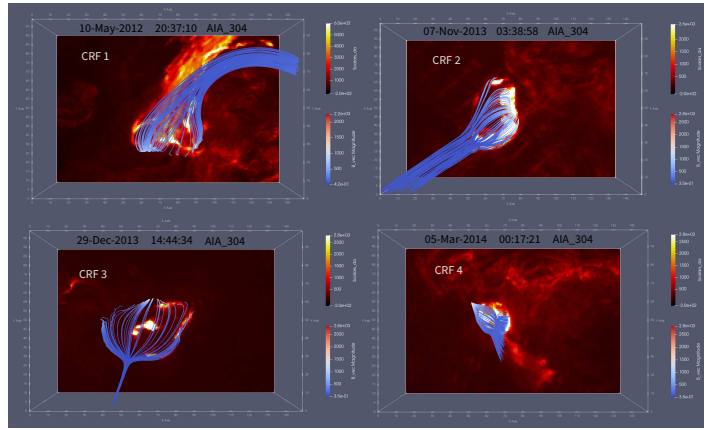}}
\caption{Nonpotential magnetic configuration of the four flares (blue lines). The bottoms of the boxes are AIA 304 {\AA} images of the flares.}
\label{fig3}
\end{figure}

The line-of-sight (LOS) and vector magnetograms of the photosphere were observed by the \textit{Helioseismic and Magnetic Imager} \citep[HMI;][]{sch12} on board SDO
with cadences of 45 s and 720 s, respectively. The HMI level\_1 data were calibrated using the SSW program \texttt{hmi\_prep.pro}.
The pre-flare vector magnetograms were used to extrapolate the coronal magnetic field.
Figure~\ref{fig2} shows vector magetograms of the four ARs hosting the flares.
We carried out potential field extrapolation based on the LOS magnetograms using the Green's function method \citep{chiu77,see78}. 
To carry out the nonlinear force-free field (NLFFF) modeling, we use vector magnetograms and the ``optimization'' method \citep{wie06,wie08}.
Pre-processing of the photospheric magnetograms is conducted before NLFFF extrapolation. 
Figure~\ref{fig3} shows the nonpotential magnetic configurations (blue lines) of the four flares. The bottoms of the boxes are EUV 304 {\AA} images of the corresponding flares.
It is clear that the magnetic configurations of the four flares are dome-like, implying the existence of well-known fan-spine topology \citep[e.g.,][]{sun13,zqm21}. 
For CRF1, the direction of possible spine line is consistent with the axis of the blowout jet (panel (a)).
Besides, the footpoints of field lines match the ribbons of the flares well, thus validating the reliability of NLFFF extrapolation.

The solar irradiance from a broad band ranging from 1$-$70 {\AA} was directly measured by the \textit{EUV SpectroPhotometer} (ESP) 
belonging to the \textit{Extreme Ultraviolet Variability Experiment} \citep[EVE;][]{wood12} on board SDO.
The \textit{Multiple EUV Grating Spectrographs} (MEGS)-A on board EVE, covering the 6$-$37 nm range, records a complete spectrum with a time cadence of 10 s and a spectra resolution of 1 {\AA}.
The standard SSW program \texttt{eve\_integrate\_line.pro} was employed to integrate irradiance over 70$-$370 {\AA} using the EVS spectral data from MEGS-A.
The soft X-ray (SXR) fluxes of the flares in 1$-$8 {\AA} were recorded by the GOES spacecraft. The isothermal temperature ($T_{e}$) and emission measure (EM) of the 
SXR-emitting plasma were derived from the ratio of GOES fluxes \citep{wh05}.
The HXR fluxes at different energy bands were obtained from the \textit{Ramaty Hight Energy Solar Spectroscopic Imager} \citep[RHESSI;][]{lin02}.
We made HXR images using the CLEAN method \citep{hur02} at energy bands of 3$-$6 and 6$-$12 keV.
The observational properties of the instruments are listed in Table~\ref{tab-2}.

\begin{table}
\caption{Description of the observational parameters.}
\label{tab-2}
\tabcolsep 1.5mm
\begin{tabular}{cccc}
\hline
Instrument & $\lambda$ & Cadence & Pixel Size\\
  & ({\AA}) & (s) & ($\arcsec$)\\ 
\hline
SDO/AIA  & 131, 171, 304 &  12  & 0.6 \\
SDO/HMI & 6173 & 45, 720 & 0.6 \\
SDO/EVE & 1$-$70 & 0.25 & ... \\
SDO/EVE & 70$-$370 & 10 & ... \\
GOES & 1$-$8 & 2.05 & ... \\
RHESSI & 3$-$50 keV & 4.0 & 4.0\\
\hline
\end{tabular}
\end{table}

\section{Energy Partition} \label{s-eng}
Using multiwavelength observations, we calculated different energy components, including: 
(i) radiative outputs in 1$-$8 {\AA}, 1$-$70 {\AA}, and 70$-$370 {\AA};
(ii) radiative loss from the SXR-emitting plasma;
(iii) peak thermal energy of the SXR-emitting plasma;
(iv) kinetic energy in flare-accelerated electrons;
and (v) magnetic free energy.

\subsection{Radiative Outputs} \label{s-rad}
As described in \citet{feng13}, the radiative output of a certain waveband ($\lambda$) is derived by integrating the background-subtracted light curve ($f_{\lambda}$),
\begin{equation} \label{eqn-1}
  U_{\lambda}=2\pi d^2\int_{t_1}^{t_2}f_{\lambda}(t)dt,
\end{equation}
where $d\approx1.496\times10^{11}$ m (1 AU) signifies the distance between the Sun and Earth, $t_1$ and $t_2$ represent the lower and upper time limits \citep{zqm19b}.

\begin{figure} 
\centerline{\includegraphics[width=1.\textwidth,clip=]{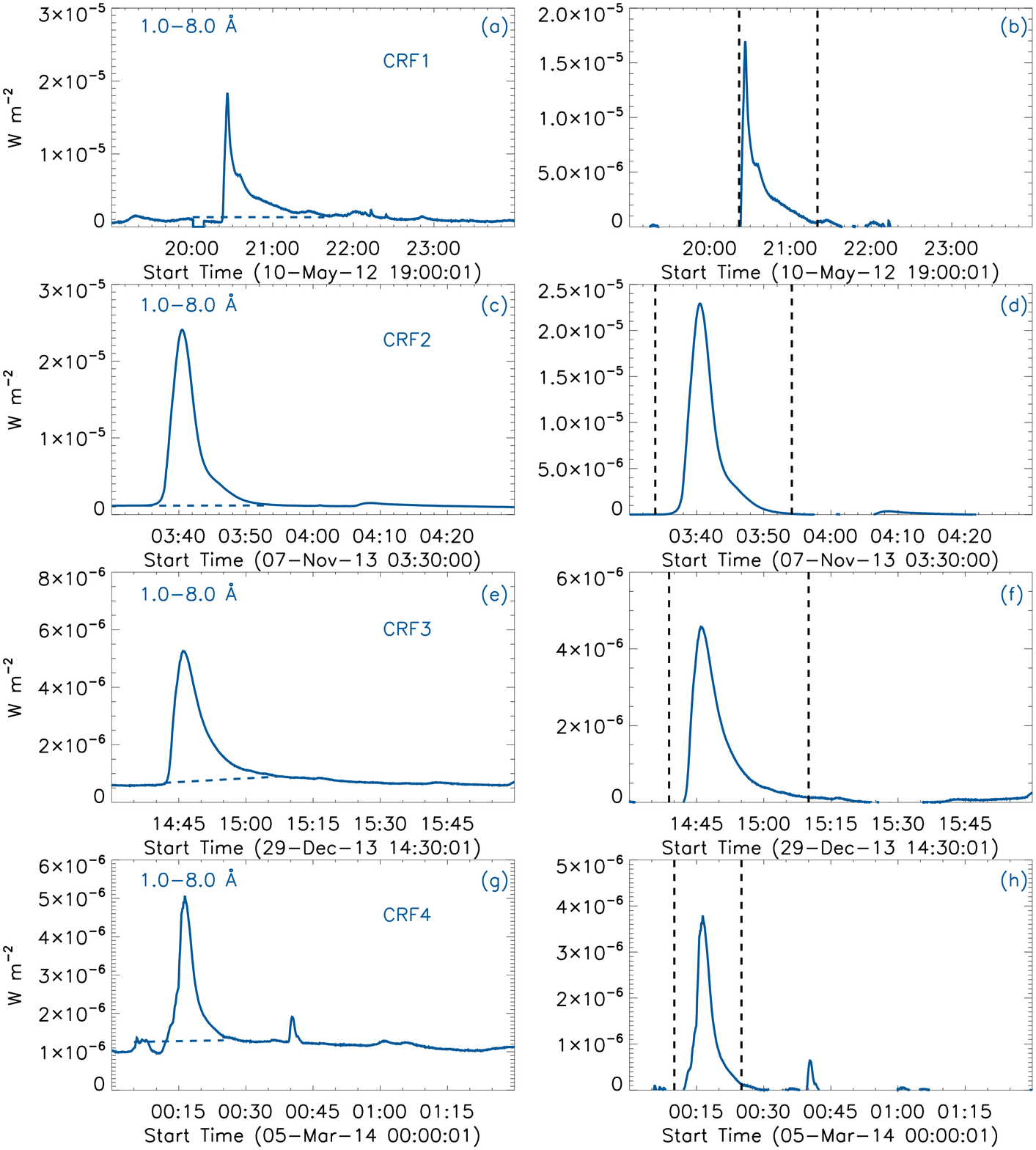}}
\caption{Left panels: SXR light curves of the flares in 1$-$8 {\AA}. The dashed lines indicate the background fluxes during the flares.
Right panels: background-subtracted light curves of the flares in 1$-$8 {\AA}. The vertical dashed lines represent the lower and upper time limits of integrals.}
\label{fig4}
\end{figure}

In Figure~\ref{fig4}, the left panels show SXR light curves of the flares in 1$-$8 {\AA}, with the dashed lines indicating the background fluxes during the flares.
The right panels show background-subtracted light curves of the flares. The radiative output $U_{1-8}$ is calculated by integrating the background-subtracted fluxes between the two dashed lines.
The values of $U_{1-8}$, falling in the range of (0.13$-$1.63)$\times$10$^{28}$ erg, are listed in the second column of Table~\ref{tab-3}.

\begin{figure}  
\centerline{\includegraphics[width=0.8\textwidth,clip=]{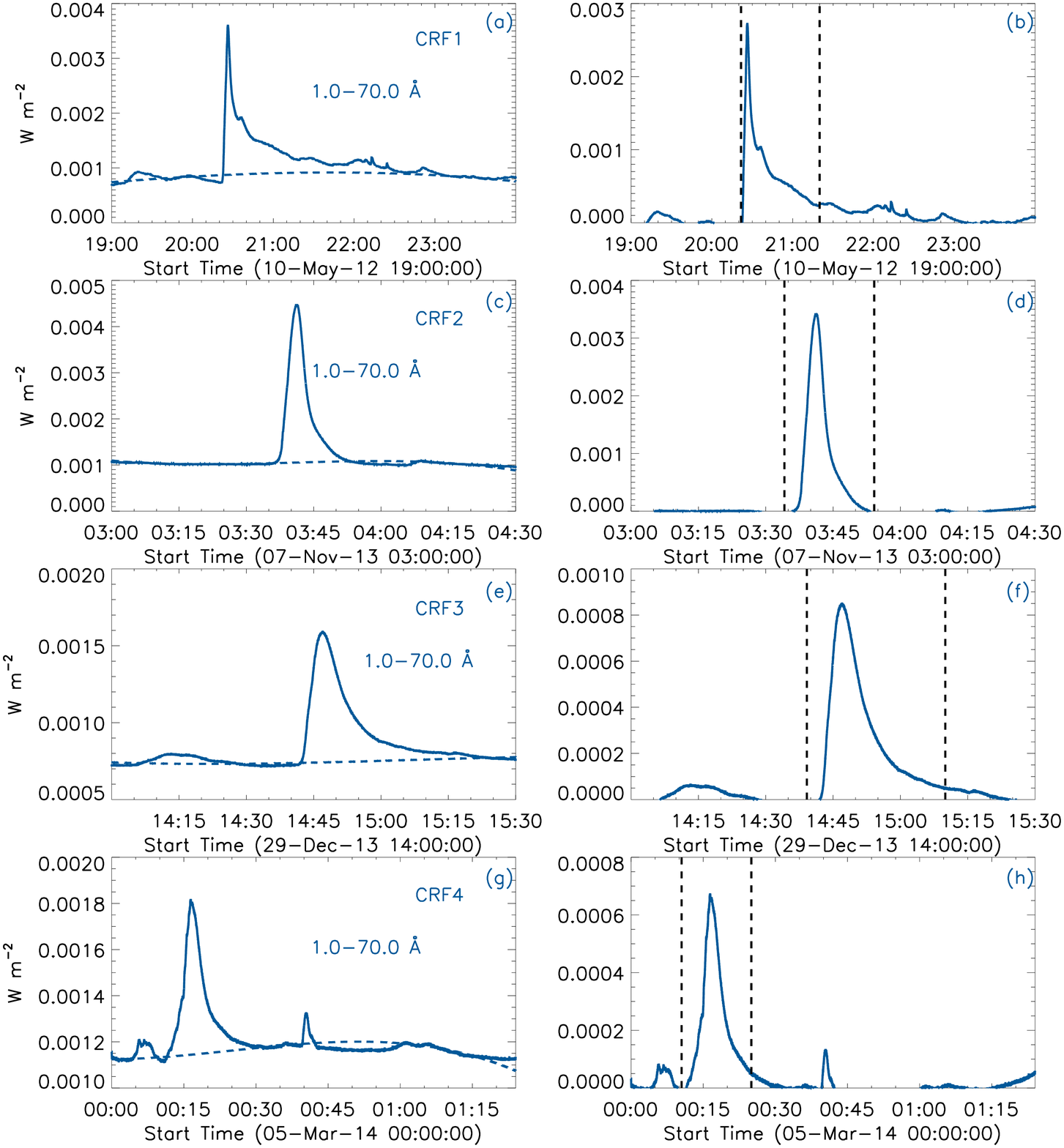}}
\caption{Left panels: light curves of the flares in 1$-$70 {\AA} observed by SDO/EVE. The dashed lines indicate the background fluxes during the flares.
Right panels: background-subtracted light curves of the flares in 1$-$70 {\AA}. The vertical dashed lines represent the lower and upper time limits of integrals.}
\label{fig5}
\end{figure}

\begin{figure} 
\centerline{\includegraphics[width=0.8\textwidth,clip=]{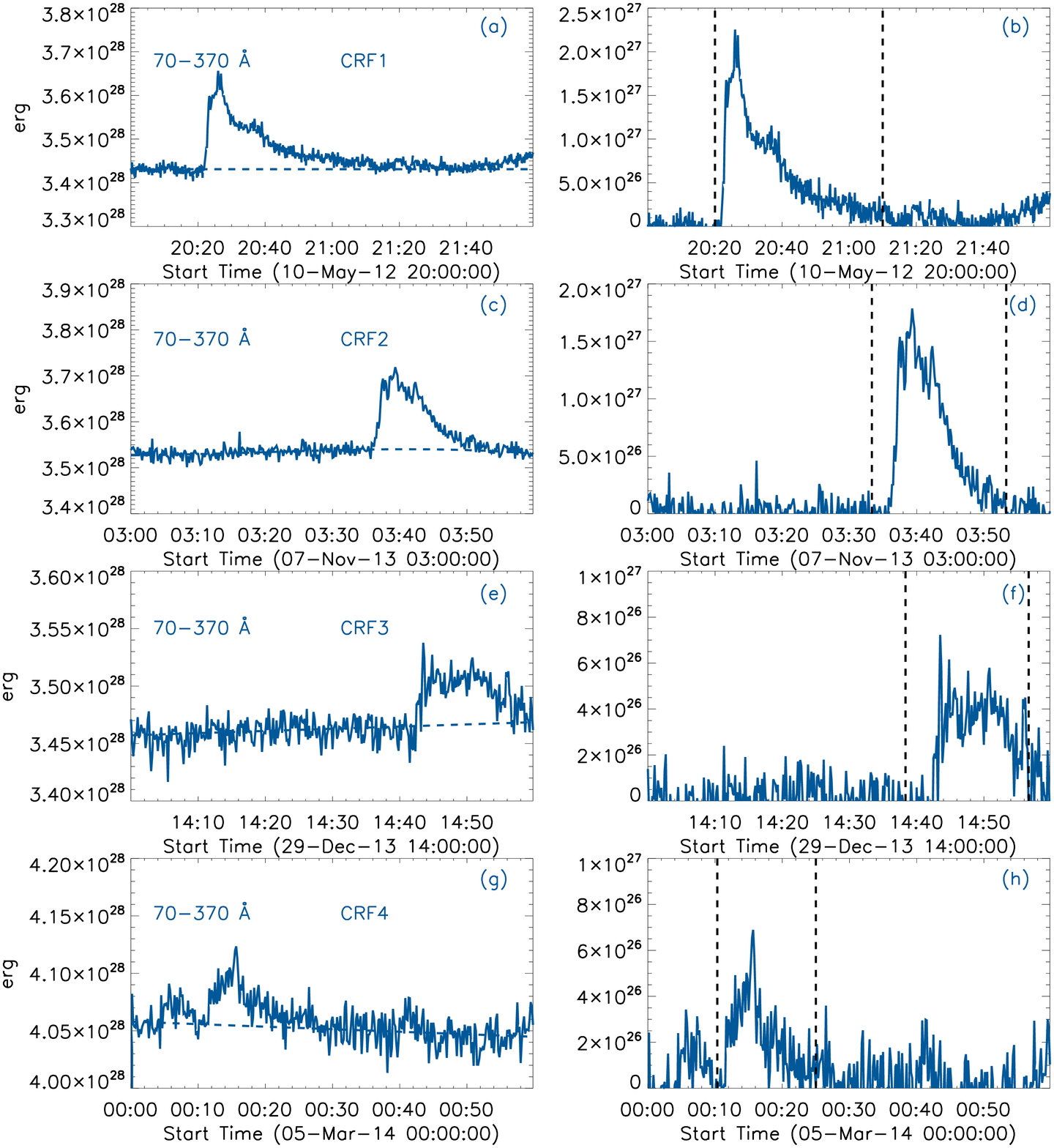}}
\caption{Left panels: light curves of the flares in 70$-$370 {\AA} observed by SDO/EVE. The dashed lines indicate the background fluxes during the flares.
Right panels: background-subtracted light curves of the flares in 70$-$370 {\AA}. The vertical dashed lines represent the lower and upper time limits of integrals.}
\label{fig6}
\end{figure}

Likewise, the left panels of Figures~\ref{fig5}-\ref{fig6} show light curves of the four flares in 1$-$70 {\AA} and 70$-$370 {\AA}.
The right panels of Figures~\ref{fig5}-\ref{fig6} show background-subtracted light curves of the flares. The radiative outputs $U_{1-70}$ and $U_{70-370}$ are calculated in the same way. 
The values of $U_{1-70}$, falling in the range of (2.8$-$41)$\times$10$^{29}$ erg, are listed in the third column of Table~\ref{tab-3}.
The values of $U_{70-370}$, falling in the range of (1.8$-$19.0)$\times$10$^{28}$ erg, are listed in the fourth column of Table~\ref{tab-3}.
It is seen that $U_{1-70}$ is 15$-$24 times larger than $U_{70-370}$ and is $\ge$200 times larger than $U_{1-8}$, which are consistent with previous results for eruptive \citep{feng13}
and confined flares \citep{zqm19b}.
The total radiative output ($U_{1-370}$) in 1$-$370 {\AA} of the flares are estimated to be the sum of $U_{1-70}$ and $U_{70-370}$, i.e., $U_{1-370}=U_{1-70}+U_{70-370}$.

\subsection{Radiative Loss from SXR-emitting Plasma} \label{s-loss}
\begin{figure} 
\centerline{\includegraphics[width=0.8\textwidth,clip=]{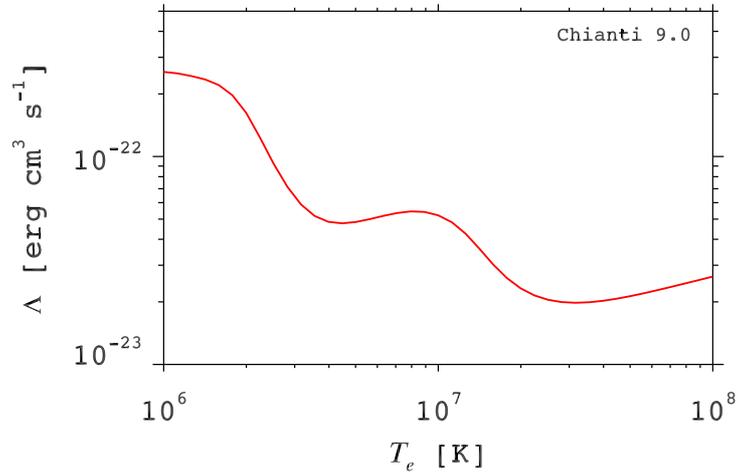}}
\caption{Radiative loss rate $\Lambda(T_e)$ as a function of temperature ($T_e$) calculated from Chianti 9.0 database.}
\label{fig7}
\end{figure}

The total radiative loss from hot plasma emitting SXR can be expressed as:
\begin{equation}  \label{eqn-2}
   T_{rad}=\int_{t_1}^{t_2}\mathrm{EM}(t)\times\Lambda(T_{e}(t))dt,
\end{equation}
where $\Lambda (T_{e})$ denotes the radiative loss rate \citep{cox69}, $\mathrm{EM}(t)$ and $T_{e}(t)$ represent the time evolution of EM and $T_e$.
Figure~\ref{fig7} shows the dependence of $\Lambda$ on $T_{e}$ in the range of 10$^6$-10$^8$ K obtained from CHIANTI 9.0 database by adopting the coronal abundances \citep{dere19}.

\begin{figure} 
\centerline{\includegraphics[width=0.8\textwidth,clip=]{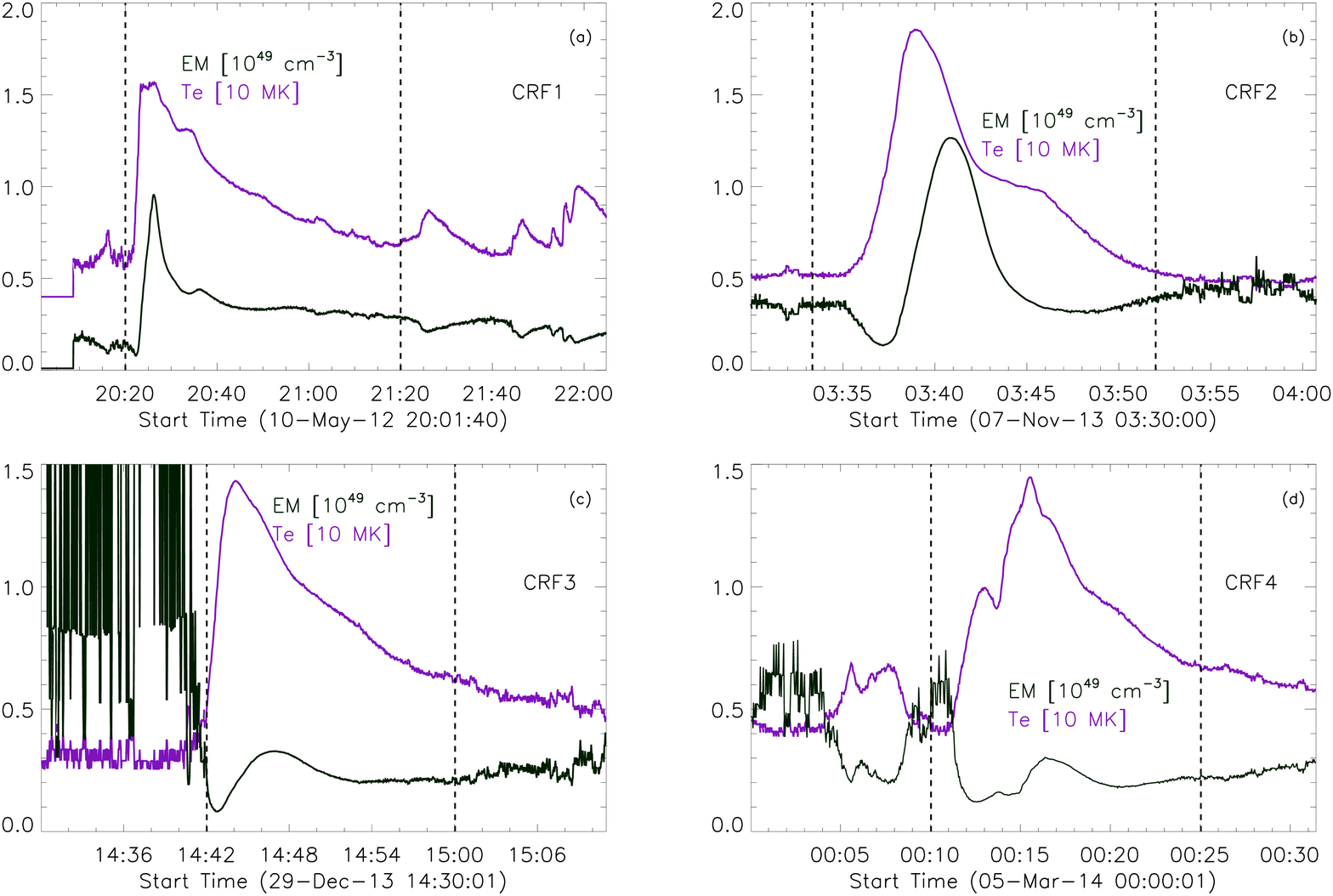}}
\caption{Time evolutions of $T_e$ and EM of the four flares obtained from GOES observations.}
\label{fig8}
\end{figure}

Figure~\ref{fig8} shows $\mathrm{EM}(t)$ and $T_{e}(t)$ of the four flares derived from GOES observations.
The vertical dashed lines indicate ${t_1}$ and ${t_2}$ for integral in Equation~\ref{eqn-2}.
The values of $T_{rad}$, being (0.75$-$2.9)$\times$10$^{29}$ erg, are listed in the fifth column of Table~\ref{tab-3}.
It is seen that $T_{rad}$ is several tens of times higher than $U_{1-8}$ \citep{feng13,zqm19b}.
    
\begin{table}
\caption{Event List with Component Energies in unit of 10$^{29}$ erg.}
\label{tab-3}
\tabcolsep 1.5mm
\begin{tabular}{ccccccccccc}
\hline
Flare & 1$-$8 & 1$-$70 & 70$-$370 & $T_{rad}$ & $E_{th,G}$ & $E_{th,R}$ & $\frac{E_{th,G}}{E_{th,R}}$ & $E_{nth}$ & $E_{mag}$ & $\frac{E_{nth}}{E_{mag}}$ \\
  \hline
CRF1 & 0.163 & 41.0 & 1.90 &  2.9 & 20.00 & 6.26 & 3.2 & 130 & 172.0 & 75.6\% \\
CRF2 & 0.085 & 17.0 & 0.73 &  2.8 & 24.60 & 4.43 & 5.6 & 69 & 97.6 & 70.7\% \\
CRF3 & 0.030 & 6.1 & 0.30 &  0.75 & 5.27 & 3.45 & 1.5 & 19 & 25.2 & 75.4\% \\
CRF4 & 0.013 & 2.8 & 0.18 &  0.86 & 5.34 & 2.88 & 1.9 & 13 & 17.6 & 73.9\% \\
\hline
\end{tabular}
\end{table}

\subsection{Peak Thermal Energy of SXR-emitting Plasma} \label{s-th}
The thermal energy of the hot plasma of flares is expressed as:
\begin{equation} \label{eqn-3}
 E_{th}=3n_{e}k_{B}T_{e}fV=3k_{B}T_{e}\sqrt{\mathrm{EM}\times fV},
\end{equation} 
where $n_{e}$ is the electron number density, $V$ is the total volume of hot plasma, and $f\approx1$ denotes the filling factor \citep{ems12,war16b}.
In the following, we calculate the peak thermal energy derived from GOES and RHESSI \citep{war16b}.

\begin{figure}
\centerline{\includegraphics[width=0.8\textwidth,clip=]{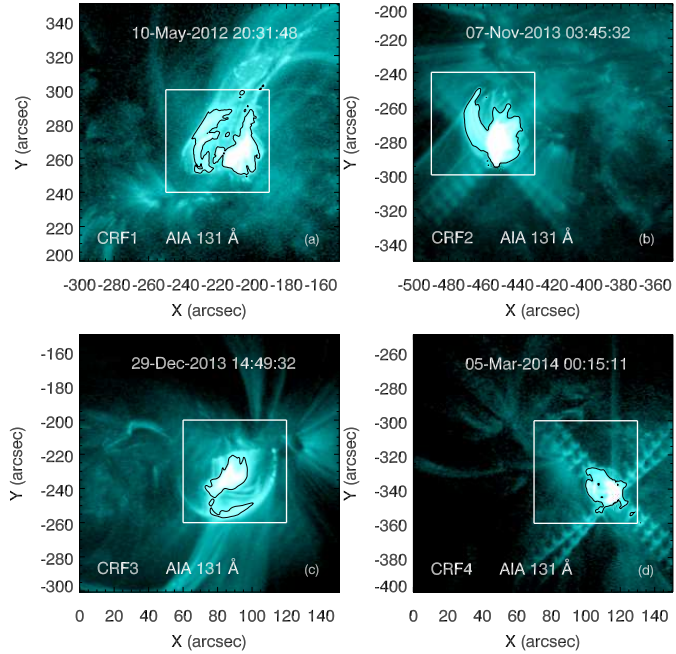}}
\caption{AIA 131 {\AA} images of the four flares near their peak times. Intensity contours of the images are drawn with black solid lines.
The white boxes indicate the areas for calculating the light curves in Figure~\ref{fig10}.}
\label{fig9}
\end{figure}

\begin{figure} 
\centerline{\includegraphics[width=0.8\textwidth,clip=]{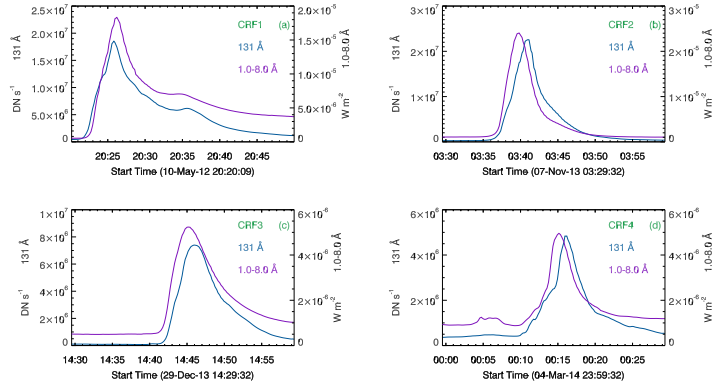}}
\caption{AIA 131 {\AA} light curves (blue lines) and SXR light curves (purple lines) of the four flares.}
\label{fig10}
\end{figure}

For CRFs, whose outer ribbons hardly expand with time, $V\approx A^{3/2}$ is assumed to be constant, where $A$ denotes the flare area encircled by the outer ribbons \citep{zqm19b}.
Figure~\ref{fig9} shows 131 {\AA} images of the four flares encircled by the white boxes when their brightness is nearly maximal.
AIA 131 {\AA} channel is dominated by the emissions of Fe\,{\sc xxi} line ($\log T\approx7.05$) during flares \citep{lem12}.
In Figure~\ref{fig10}, the 131 {\AA} light curves of the flares (blue lines) are compared with the SXR light curves (purple lines), 
showing that the light curves have good correlations with correlation coefficients of 0.96, 0.80, 0.94, and 0.90, respectively.
Therefore, the hot plasma observed in 131 {\AA} serves as a proxy of SXR-emitting plasma.
The areas of flares are calculated by summing up the pixels whose intensities are above an ad hoc criterion,
which is taken to be $\sim$20 times higher than the average intensity of the nearby quiet region.
The projection effect of $A$ is corrected by multiplying a factor of $(\cos \mu)^{-1}$, where $\mu$ signifies the longitude of flare core. 
The corresponding $A$ and $V$ in 131 {\AA} are listed in the second and fourth columns of Table~\ref{tab-4}.
Combining the four flares in this study with the two M1.1 flares in AR 12434, 
it is found that the thermal source volumes are systematically larger in M-class flares than C-class flares \citep{war20}.

\begin{table}
\caption{Evaluation of the area ($\times$10$^{18}$ cm$^2$), volume ($\times$10$^{28}$ cm$^3$), 
and peak thermal energy ($\times$10$^{29}$ erg) of hot plasma.}
\label{tab-4}
\tabcolsep 1.5mm
\begin{tabular}{ccccccc}
\hline
Flare & $A_{\mathrm{131}}$ & $A_{\mathrm{HXR}}$ & $V_{\mathrm{131}}$ & $V_{\mathrm{HXR}}$ & $E_{th,G}$ & $E_{th,R}$ \\
  \hline
CRF1 & 4.63 & 2.69 & 1.00 & 0.44 & 20.00 & 6.26 \\
CRF2 & 4.86 & 1.78 & 1.07 & 0.24 & 24.60 & 4.43 \\
CRF3 & 2.11 & 2.46 & 0.31 & 0.39 & 5.27 & 3.45 \\
CRF4 & 2.16 & 2.36 & 0.32 & 0.36 & 5.34 & 2.88 \\
\hline
\end{tabular}
\end{table}

Equation~\ref{eqn-3} indicates that the peak thermal energy is reached when $T_{e}\sqrt{\mathrm{EM}}$ is maximal.
Using observations from GOES (Figure~\ref{fig8}), we calculated the peak values $E_{th,G}$ of the flares, which are listed in Table~\ref{tab-3} and Table~\ref{tab-4}.
The peak values $E_{th,G}$ fall in the range of (5.3$-$24.6)$\times$10$^{29}$ erg.
The total heating requirements of the flares, including the peak thermal energy and radiative loss, are estimated to be (0.6$-$2.7)$\times$10$^{30}$ erg.
Conductive energy loss is not considered in this study since conduction may be severely suppressed or conduction loss is recycled through conduction-driven evaporation \citep{war20}. 
Note that CRF1 was accompanied by a blowout jet. The total thermal energy of CRF1 and the jet is estimated to be (2.3$-$2.5)$\times$10$^{30}$ erg, 
considering that the thermal energies of jets account for $\frac{1}{7}$ to $\frac{1}{4}$ of the footpoint flares \citep{shi00}.
       
\begin{figure} 
\centerline{\includegraphics[width=0.8\textwidth,clip=]{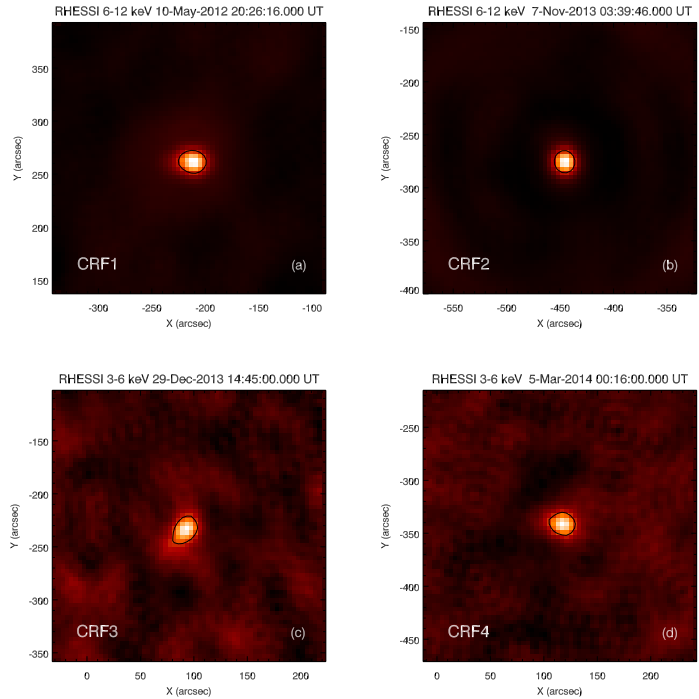}}
\caption{HXR images of the four flares near the HXR peak times. 
The energy bands are 6$-$12 keV for the two M-class flares and 3$-$6 keV for the two C-class flares, respectively.
The black lines represent contours of 50\% peak intensities.}
\label{fig11}
\end{figure}

Figure~\ref{fig11} shows the HXR images of the four flares near the HXR peak times.
The energy bands are 6$-$12 keV for the two M-class flares and 3$-$6 keV for the two C-class flares.
It is clear that the HXR source is single and small for each flare. The contours of 50\% peak intensities are drawn with black lines.
The 6-12 keV sources of the two M-class flares and 3-6 keV sources of the two C-class flares are considered.
The area of thermal source observed by RHESSI is taken to be the total area of pixels within the black lines.
However, we found that the area and volume are comparable for different energy bands.
The corresponding values of $A$ and $V$ are listed in the third and fifth columns of Table~\ref{tab-4}.

Figure~\ref{fig12} shows selected HXR spectra of the four flares obtained from RHESSI observations.
The spectra are fitted with a combination of a thermal component and a thick-target nonthermal component.
The fitting is performed using the standard SSW program \texttt{thick2.pro} in the \texttt{OSPEX} package.
The parameters of thermal component, including $T$ in unit of MK and EM in unit of 10$^{49}$ cm$^{-3}$, are labeled.
Using Equation~\ref{eqn-3}, the peak thermal energies ($E_{th,R}$) derived from RHESSI are calculated and listed in Table~\ref{tab-3} and Table~\ref{tab-4}.
The ratio of $\frac{E_{th,G}}{E_{th,R}}$ is accordingly obtained and listed in the eighth column of Table~\ref{tab-3}. 
It is revealed that the ratio is greater than 1.0 for all events, which is consistent with previous results \citep{war20}.

\subsection{Nonthermal Energy in Flare-accelerated Electrons} \label{s-non}
\begin{figure}
\centerline{\includegraphics[width=0.8\textwidth,clip=]{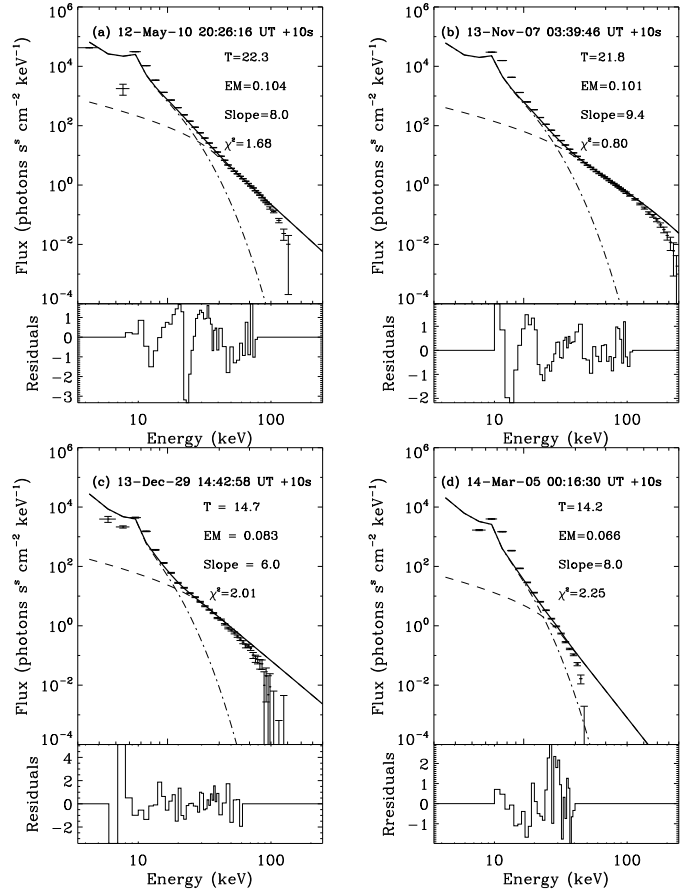}}
\caption{Selected HXR spectra of the four flares obtained from RHESSI observations and the corresponding normalized residuals of the spectral fitting.
The observed data are represented by the points with error bars.
The fitted thermal and nonthermal power-law components are drawn with dot-dashed lines and dashed lines, respectively.
The sum of both components are drawn with thick solid lines.
The fitted parameters, including $T$ in unit of MK, EM in unit of 10$^{49}$ cm$^{-3}$, and power-law index of electrons are labeled.}
\label{fig12}
\end{figure}

To estimate the nonthermal energy in flare-accelerated electrons, we integrated the power of injected electrons over time \citep{ning10}:
\begin{equation} \label{eqn-4}
   E_{nth}=\int_{t_1}^{t_2}P_{nth}(t)dt=\int_{t_1}^{t_2}\frac{dE_{nth}}{dt}dt,
\end{equation}
where $t_1$ and $t_2$ represent the start and end times of flare at energy band of 25$-$50 keV.
Here, $P_{nth}(t)$ can be calculated by integrating the electron power-law spectrum above a low-energy cutoff ($E_c$) and below a high-energy cutoff ($E_{h}\approx30$ MeV):
\begin{equation} \label{eqn-5}
   P_{nth}(t)=\frac{dE_{nth}}{dt}=\int_{E_c}^{E_h}A_{0}E_{0}^{-\delta}dE_{0},
\end{equation}
where $A_0$ is the electron flux in unit of 10$^{35}$ electrons s$^{-1}$, and $\delta$ is the power-law index of nonthermal electrons (see Figure~\ref{fig12}).
The values of $E_c$ are 20, 30, 23, and 28 keV, respectively. Using Equation~\ref{eqn-4} and the above parameters, we estimated the total nonthermal energy in flare-accelerated electrons.
The values of $E_{nth}$, falling in the range of (1.3$-$13)$\times$10$^{30}$ erg, are listed in the ninth column of Table~\ref{tab-3}.
The ratio of $E_{nth}/E_{th,G}$ is between $\sim$2.4 and $\sim$6.5.
It should be emphasized that these calculated values are lower limits of real nonthermal energies, which rely sensitively on $E_{c}$ \citep{war20}. 
Besides, we did not consider the nonthermal energy in flare-accelerated ions.

\subsection{Magnetic Free Energy} \label{s-mag}
As mentioned in Section~\ref{s-obs}, Figure~\ref{fig2} shows the vector magnetograms of the four ARs where flares took place. Both potential field and nonpotential field extrapolations were performed.
Figure~\ref{fig3} shows the nonpotential magnetic field lines (blue lines) of the flares.
The magnetic free energy ($E_{mag}$) is defined as the excess magnetic energy of the NLFFF ($E_{np}$) relative to the energy of potential field ($E_{p}$):
\begin{equation} \label{eqn-6}
 E_{mag}=E_{np}-E_{p}=\int_{V}\frac{B_{np}^2-B_{p}^2}{8\pi}dV.
\end{equation}
We calculated $E_{mag}$ in the flare regions as enclosed by the red boxes in Figure~\ref{fig2}.
The estimated $E_{mag}$, ranging from 1.8$\times$10$^{30}$ to 1.7$\times$10$^{31}$ erg, are listed in the tenth column of Table~\ref{tab-3}.
It is obvious that the free magnetic energies are larger than the nonthermal energies and radiative output in 1$-$370 {\AA}, indicating that the accumulated free energy before flare 
is sufficient to provide the kinetic energy in flare-accelerated energies and radiation, thus validating the magnetic nature of confined flares \citep{pri02}.
The ratio of $E_{nth}/E_{mag}$ for CRFs falls in the range of 70\%$-$76\% (see the last column of Table~\ref{tab-3}), 
which is much higher than that of X-class eruptive flares \citep{feng13,tha15}. 
In other words, more free energy is converted into the kinetic energy of flare-accelerated electrons in confined flares than in eruptive flares, 
because a large fraction of free energy is converted into the kinetic, thermal, and potential energies of CMEs for eruptive flares \citep{ree10,ems12,feng13}.

\section{Discussion} \label{s-dis}
\begin{figure}
\centerline{\includegraphics[width=0.8\textwidth,clip=]{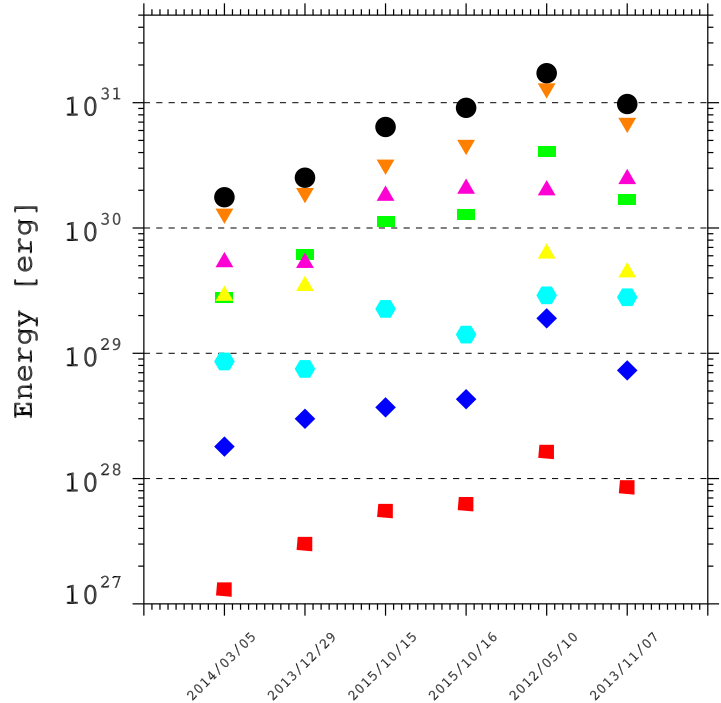}}
\caption{Energy components of the four events in this study and the previous two events in \citet{zqm19b}.
The six events are arranged with increasing flare importance (see text for detail).
The radiative outputs in 1$-$8 {\AA}, 1$-$70 {\AA}, 70$-$370 {\AA}, total radiative loss, peak thermal energy derived from GOES and RHESSI, nonthermal energy in electrons, 
and magnetic free energy are labeled with red squares, green rectangles, blue diamonds, cyan hexagons, magenta triangles, yellow triangles, orange triangles, and black circles, respectively.}
\label{fig13}
\end{figure}

As mentioned in Section~\ref{s-intro}, \citet{zqm19b} explored the energy partition in two M-class CRFs. 
The radiative outputs in 1$-$8 {\AA} and 1$-$70 {\AA} are obtained using the observations of GOES and SDO/EVE.
Total radiative loss and peak thermal plasma are calculated using the observations of GOES and SDO/AIA.
Nonthermal energy of electrons are derived using the observation of RHESSI (see their Table 2).
The radiation in 70$-$370 {\AA}, total solar irradiance, nonthermal energy of ions, and dissipated magnetic free energy are estimated according to previous statistical works.
In this study, we calculated the radiation in 70$-$370 {\AA} using the observations of SDO/EVE and 
magnetic free energy using magnetic extrapolation based on the vector magnetograms from SDO/HMI.
The results combining these six events with increasing flare importance (or peak GOES flux) are plotted in Figure~\ref{fig13}.
The orders of magnitude of the energy components are clearly demonstrated. 
Moreover, for each component, the values increase systematically with flare importance, suggesting that more energy is involved in larger flares.
Our findings are in accordance with previous statistical results \citep{war20}.

\begin{figure}
\centerline{\includegraphics[width=0.8\textwidth,clip=]{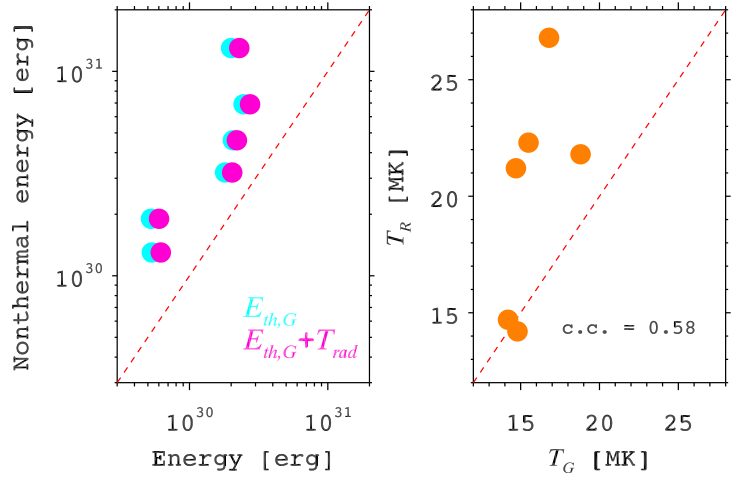}}
\caption{Left panel: scatter plot showing the relationship between nonthermal energy and thermal energy (cyan circles) and heating requirement (magenta circles).
Right panel: scatter plot showing the relationship between the maximal temperatures of GOES ($T_G$) and RHESSI ($T_R$). The correlation coefficient ($\sim$0.58) is labeled.
The red dashed lines in both panels represent the $y=x$ curves.}
\label{fig14}
\end{figure}

In Figure~\ref{fig14}, the left panel shows the scatter plot of the six events. The relationship between the nonthermal and thermal energies is illustrated with cyan circles,
while the relationship between the nonthermal energy and heating requirement (including the thermal energy and radiative loss) is illustrated with magenta circles.
There is a good linear correlation between the nonthermal energy and thermal energy, validating the previous results for nine medium-sized flares \citep{sai05}.
It is obvious that the nonthermal energies are higher than the heating requirements of hot plasma, at least for the six events we have studied.
\citet{kush15} investigated an M6.2 confined flare on 2004 July 14. 
The peak thermal energy and nonthermal energy are calculated to be 3.89$\times$10$^{29}$ erg and 3.03$\times$10$^{30}$ erg, respectively.
Hence, the ratio of $E_{nth}/E_{th}$ reaches $\sim$7.5, which is consistent with the present work.

The right panel of Figure~\ref{fig14} shows the scatter plot of the six events to illustrate the relationship between the maximal temperatures of GOES ($T_G$ in MK) and RHESSI ($T_R $ in MK).
It is seen that $T_R$ is higher than $T_G$ in most cases and a good linear correlation exists between the two parameters. A linear fit yields $T_R=1.65T_G-5.85$, which lies between 
$T_R=1.12T_G-3.12$ \citep{bat05} and $T_R=1.78T_G-4.61$ \citep{war16a}. It is noted that our study has limitations due to the small sample size. 
Additional statistical studies using more events and numerical simulations are worthwhile to draw a decisive conclusion.

\section{Summary} \label{s-sum}
In this paper, we investigated the energy partition of four confined circular-ribbon flares near the solar disk center. 
Using multiwavelength observations from SDO, GOES, and RHESSI, we calculated different energy components, including the radiative outputs in 1$-$8, 1$-$70, and 70$-$370 {\AA},
total radiative loss, peak thermal energy derived from GOES and RHESSI, nonthermal energy in flare-accelerated electrons, and magnetic free energy before flares.
The main results are as follows:
\begin{enumerate}
\item The energy components increase systematically with the flare importance or peak GOES flux, indicating that more energies are involved in larger flares. 
The magnetic free energies are larger than the nonthermal energies and radiative outputs of flares, which is consistent with the magnetic nature of flares. 
The ratio $\frac{E_{nth}}{E_{mag}}$ of the four flares, being 0.70$-$0.76, is considerably higher than that of eruptive flares. 
Hence, this ratio may serve as an important factor that discriminates confined and eruptive flares.
The nonthermal energies are sufficient to provide the heating requirements including the peak thermal energy and radiative loss. 
\item Our findings impose constraint on theoretical models of confined CRFs and have potential implication for the space weather forecast. 
Statistical studies based on a larger sample are especially needed to draw decisive conclusions.
\end{enumerate}

\begin{acknowledgements}
The authors are grateful to the referee for valuable suggestions.
The authors thank Drs. Tie Liu, Yanjie, Liu, and Ya Wang for helpful discussion.
SDO is a mission of NASA\rq{}s Living With a Star Program. AIA and HMI data are courtesy of the NASA/SDO science teams. 
This work is funded by NSFC grants (No. 11790302, 11773079, 41761134088, 11473071), the International Cooperation and Interchange Program (11961131002), 
the Youth Innovation Promotion Association CAS, CAS Key Laboratory of Solar Activity, National Astronomical Observatories (KLSA202006),
and the Strategic Priority Research Program on Space Science, CAS (XDA15052200, XDA15320301).
\end{acknowledgements}

\end{article}
\end{document}